\documentclass[preprint,12pt]{elsarticle}
\usepackage{amssymb}
\usepackage{amsthm}
\usepackage{amsfonts}
\usepackage{subeqnarray}
\usepackage[latin1]{inputenc}

\journal{arXiv.org}

\begin{document}

\begin{frontmatter}

\title{Fracture Propagation Driven by Fluid Outflow \\ from a Low-permeability Aquifer}

\author[label1]{Gennady Y. Gor}
\ead{ggor@princeton.edu}
\author[label2]{Howard A. Stone}
\author[label1]{Jean H. Pr\'evost}

\address[label1]{Department of Civil and Environmental Engineering, Princeton University, \\ Princeton, New Jersey 08544, United States}
\address[label2]{Department of Mechanical and Aerospace Engineering, Princeton University, \\ Princeton, New Jersey
08544, United States}

\fnref{label3}
\fntext[label3]{Corresponding author, Tel.: +1 (609) 258-1619, Fax: +1 (609) 258-2760}

\begin{abstract}

Deep saline aquifers are promising geological reservoirs for CO$_2$ sequestration if they do not leak. The absence of leakage is provided by the caprock integrity. However, CO$_2$ injection operations may change the geomechanical stresses and cause fracturing of the caprock. We present a model for the propagation of a fracture in the caprock driven by the outflow of fluid from a low-permeability aquifer. We show that to describe the fracture propagation, it is necessary to solve the pressure diffusion problem in the aquifer. We solve the problem numerically for the two-dimensional domain and show that, after a relatively short time, the solution is close to that of one-dimensional problem, which can be solved analytically. We use the relations derived in the hydraulic fracture literature to relate the the width of the fracture to its length and the flux into it, which allows us to obtain an analytical expression for the fracture length as a function of time. Using these results we predict the propagation of a hypothetical fracture at the In~Salah CO$_2$ injection site to be as fast as a typical hydraulic fracture. We also show that the hydrostatic and geostatic effects cause the increase of the driving force for the fracture propagation and, therefore, our solution serves as an estimate from below. Numerical estimates show that if a fracture appears, it is likely that it will become a pathway for CO$_2$ leakage.

\end{abstract}

\end{frontmatter}

\section{Introduction}
\label{intro}

Use of fossil fuels for satisfaction of current energy needs has an inherent waste product -- carbon dioxide. Since the beginning of the technological revolution the amount of CO$_2$ released in the atmosphere has grown monotonically, causing a substantial increase of its concentration. Within the last decade, a significant effort has been expended on identifying ways to avoid CO$_2$ release in the atmosphere, which is the domain of CO$_2$ sequestration. Various geological formations are considered as options for long-term storage of CO$_2$: depleted oil reservoirs, unmineable coal seams, deep saline aquifers, etc. The latter are especially promising because they are widespread and have high capacity. 

Deep aquifers are separated from the shallow freshwater aquifers by caprock -- a formation with extremely low permeability (often shale). When CO$_2$ is injected into an aquifer, the integrity of the caprock prevents CO$_2$ leakage. However, buildup of the fluid pressure caused by CO$_2$ injection changes the stresses in the caprock, and can lead to reactivation of preexisting faults \cite{Cappa-Rutqvist} or even fracturing of the caprock \cite{Iding-2010}.

Recent studies have shown that when CO$_2$ is injected at a temperature lower than the ambient temperature of the formation, additional thermal stresses develop around the injection well and the risk of fracturing increases \cite{Luo-2010}, \cite{Luo-2011}, so that even the caprock can be fractured \cite{Preisig-Prevost}, \cite{Goodarzi-2012}, \cite{Gor-thermal}. In our recent work \cite{Gor-thermal} we revealed two regions of high tensile stresses, where fracturing may occur: (1) in the immediate vicinity of the injection well, and (2) above the injection well in the caprock at the boundary with the aquifer. The first can lead to horizontal fractures in the aquifer, which are of no concern (and are even beneficial, since they can increase injectivity). The second can lead to short vertical fractures in the caprock. However, fracturing does not necessarily lead to leakage; CO$_2$ will leak out of the aquifer only if the fractures are long enough to reach an abandoned well \cite{Humez-2011} or connect to a network of natural fractures \cite{Smith-2011}. The initial length of the fractures can be small, e.g. of the order of 10~cm to 10~m \cite{Gor-thermal}, but under high fluid pressure the fractures may propagate. Therefore, the rate of fracture propagation and the characteristic length of fractures are crucial for assessing the possibility of CO$_2$ leakage from a deep aquifer.

Fluid-driven fracture propagation involves multiple physical processes: fracture mechanics, flow in the fracture and flow in the porous aquifer. However, when an aquifer has low permeability, the fluid outflow from it is slow and therefore it is the rate-limiting process for fracture propagation. Therefore, in order to predict the rate of fracture propagation, one has to calculate the outflow (discharge) from the aquifer, which can be found from the solution of the pressure equation.

The evolution of pressure takes place in two regions: the aquifer and the propagating fracture. However, we show here that when the permeability of the aquifer is significantly lower than the permeability in the fracture then it can be assumed that the pressure in the fracture is established instantaneously. This assumption is used in the hydraulic fracture literature \cite{Geertsma-deKlerk}. Therefore, the pressure diffusion problem can be considered only in the aquifer. 

Even when considering the pressure diffusion only in the aquifer, the problem is non-trivial, since it is an unsteady problem in a two-dimensional (2D) domain. We solve the 2D problem numerically and find that after a relatively short time, the solution for the flux is equal to twice the solution of a simplified one-dimensional (1D) problem from the two horizontal flow paths toward the fracture. The 1D problem can be solved analytically. 

The analytical solution for the pressure diffusion problem provides an expression for the fluid flux into the fracture. Then, assuming the Khristianovich-Geertsma-de Klerk (KGD) geometry \cite{Khr-Zheltov} for the fracture and using the relations for the fracture aperture from \cite{Geertsma-deKlerk}, from the calculated flux we can obtain the fracture length and aperture as a function of time. 

Using our analytical solution we make estimates based on the parameters for the Krechba aquifer (In Salah, Algeria) from \cite{Preisig-Prevost}, \cite{Rutqvist-2010}. This site is of significant technological interest because it has been used as a pilot project for CO$_2$ injection since 2004. We find that initially the fracture propagation is very fast, similar to the rate of propagation of hydraulic fractures. Our analytical solution predicts fracture propagation of $100$ meters within less then a minute after initiation. On such length scales a fracture may easily reach a leaky fault, a system of natural fractures or an abandoned well and become a pathway for CO$_2$ leakage from the aquifer into potable aquifers or even into the atmosphere. We also show that the hydrostatic and geostatic effects cause the increase of the driving force for the fracture propagation and, therefore, our solution serves as an estimate from below. 

\section{Problem Formulation and Model}
\label{model}

\begin{figure}[h]
  \centering
  \includegraphics[width=1.0\linewidth]{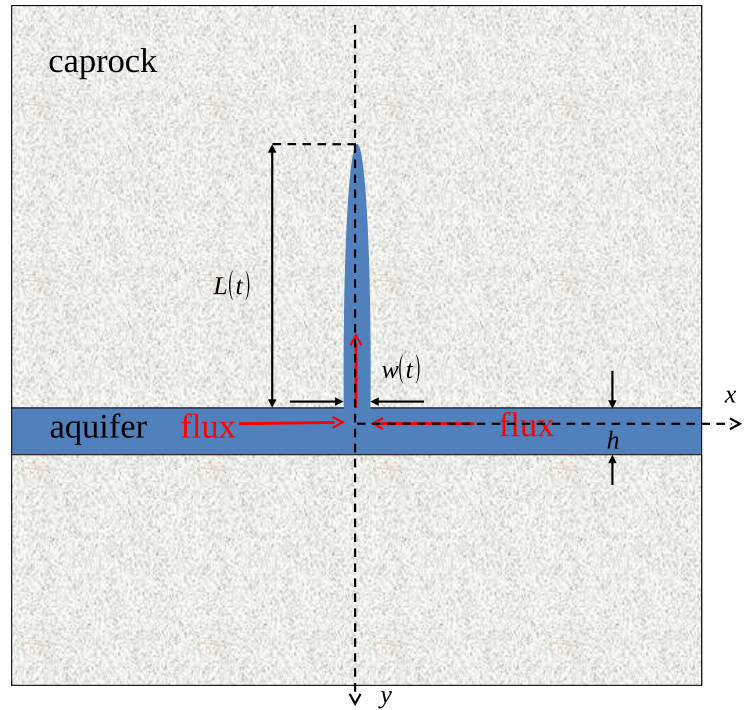}  
  \caption{System of an aquifer and caprock with a vertical elliptic fracture: two-dimensional representation. The fluid flux from the over-pressured aquifer (from $x = -\infty$ and $x = + \infty$) drives the fracture propagation.}
  \label{fig:scheme}
\end{figure}

We consider the physical system to consist of a porous aquifer filled with fluid (brine and injected supercritical carbon dioxide) and the caprock (shale) that constrains the aquifer from above. We assume that the aquifer has relatively low permeability ($\sim 10-100$ mD), which is the case, for example, for the sandstone aquifer at the Krechba field (In Salah, Algeria). Injection of cold CO$_2$ leads to a pressure buildup in the aquifer and to tensile stresses in the caprock. Our recent simulations \cite{Gor-thermal} showed that after several years of continuous injection of cold CO$_2$ the stresses in the caprock above the horizontal injection well exceed the tensile strength of the caprock. Therefore, the caprock fractures. Here we do not discuss the evolution of stresses and initiation of the fracture, since that has been done in ref. \cite{Gor-thermal}. Rather, we consider a single vertical 2D fracture originating at the boundary between the 2D aquifer and the caprock, and we assume that the fracture has an elliptical KGD geometry \cite{Khr-Zheltov}, \cite{Barenblatt-1962}, \cite{Geertsma-deKlerk}; a schematic of the system is represented in Figure \ref{fig:scheme}. 

High pressure in the aquifer pushes the fluid into the fracture, which may cause it to propagate further. There are several physical mechanisms controlling the behavior of a fluid-driven fracture. For a typical well-driven hydraulic fracturing operation, the injected flow rate is high and the fracture propagation rate is limited by two dissipative processes: fracturing of the rock (controlled by the rock toughness) and dissipation in the fluid (controlled by fluid viscosity) \cite{Detournay-2004}. However, the case considered here differs substantially. The source of fluid is the aquifer, which has low permeability, and therefore the outflow of fluid from it is relatively slow. The rate of fracture propagation cannot be faster than the flow of fluid that causes this propagation. Since the fluid outflow from the aquifer is the rate-limiting process for the fracture propagation, it is the only process considered below. 

If a fluid-driven fracture propagates in a permeable media, the fluid may seep into the rock through the walls of the fracture. When the permeability of the rock is high, this effect may noticeably affect the rate of propagation \cite{Nilson-1988}, but in our case a fracture propagates in shale with typical permeabilities of the order of $10^{-6}$ mD \cite{Zhang-Scherer}, so the leak-off effects can be neglected.

We denote the fracture length $L$ and the aperture (maximum width) $w$, which are both functions of time $t$ when the fracture propagates. The time $t=0$ corresponds in our model to the initiation of the fracture, which (according to \cite{Gor-thermal}) may take place after several years of continuous injection of CO$_2$. We assume that the initial length of the fracture $L(0)$ is negligibly small compared to its length as it propagates.

We denote by $y$ the direction into the aquifer, so that $-y$ is the direction of fracture propagation and $y=0$ denotes the interface between aquifer and caprock. Following \cite{Geertsma-deKlerk} we assume that the fracture is filled with fluid, which originated in the aquifer and flowed into the fracture; the pressure in the fracture equilibrates instantaneously. Figure \ref{fig:scheme} represents the system under consideration.

We begin with the material balance equation: the rate of change of volume of the fracture $V$ is equal to the fluid volumetric flow rate brought from the aquifer, i.e. discharge $Q$,
\begin{equation}
\label{eq:balance}
\frac{dV(t)}{dt} = Q(t).
\end{equation}
The main goal for us is, therefore, to calculate the flux $Q(t)$. Once we know it, we can predict how the fracture volume and length change. 

The flux $Q$ can be found from the evolution of pressure $p(x,y,t)$ in the 2D aquifer with leakage through the opening; $x$ is the horizontal axis, the $y$ axis is positive downwards, $x=0$, $y=0$ correspond to the center of the fracture opening, and the $z$ axis is perpendicular to the plane of the image in Figure \ref{fig:scheme}. The pressure evolution is governed by a diffusion equation \cite{Coussy}
\begin{equation}
\label{eq:diff2d}
\frac{\partial p(x,y,t)}{\partial t} = c_f \left( \frac{\partial^2 p(x,y,t)}{\partial x^2} + \frac{\partial^2 p(x,y,t)}{\partial y^2} \right)
\end{equation}
where 
\begin{equation}
\label{eq:c_f}
c_f = \frac{k}{\mu \phi C}
\end{equation}
is the diffusion coefficient for pressure, $\phi$ is the porosity of the aquifer, and $C$ is the compressibility of the fluid. Characteristic values of $\phi$, $k$, $\mu$ and $C$ (see Table \ref{tab:param}) give $c_f \simeq 0.2$ m$^2$/s.

\begin{table}
\caption{Properties of In Salah site for CO$_2$ injection}
\begin{center}
\begin{tabular}{p{8cm}p{4cm}}
\hline
Permeability of aquifer$^*$ $k$ & $50$~mD \\
Permeability of caprock$^{\dagger}$ $k_s$ & $10^{-6} - 10^{-4}$~mD \\
Porosity of aquifer$^{\dagger}$  $\phi$ & $0.17$ \\
Porosity of caprock$^{\dagger}$ $\phi_s$ & $0.01$ \\
aquifer temperature$^*$ $T$ & $90^{\circ}$C \\
CO$_2$ injection temperature$^*$ $T_i$ & $50^{\circ}$C \\
Density of CO$_2^*$ $\rho_{CO_2}$ & $900$~kg/m$^3$ \\
Density of brine$^{\mathsection}$ $\rho_w$ & $1000$~kg/m$^3$ \\
Density of caprock$^*$ $\rho_s$ & $2400$~kg/m$^3$ \\
Compressibility of CO$_2^*$ & $1.3\times10^{-8}$~Pa$^{-1}$ \\
Compressibility of brine$^{\mathsection}$ & $4.1\times10^{-10}$~Pa$^{-1}$ \\
Viscosity of CO$_2^*$ & $9.0\times10^{-5}$~Pa$\cdot$s \\
Viscosity of brine$^{\mathsection}$ & $3.0\times10^{-4}$~Pa$\cdot$s \\
Fluid pressure in aquifer$^*$ $p_0$ & $30$~MPa \\
Confining horizontal stress$^*$ $\sigma$ & $28$~MPa \\
Water residual saturation$^*$ $S_{rw}$ & 0.25 \\
Diffusion coefficient for pressure $c_f$ & $0.215$~m$^2$/s \\
Young's modulus of the caprock$^{\dagger}$ $E$ & $20$~GPa \\
Poisson's ratio of the caprock$^{\dagger}$ $\nu$ & $0.15$ \\
\hline
\end{tabular}
\begin{tabular}{p{12cm}}
$^*$ from \cite{Gor-thermal} \\
$^{\dagger}$ from \cite{Rutqvist-2010} and \cite{Preisig-Prevost} \\
$^{\mathsection}$ from \cite{Gor-thermal}, values for water are used \\
\hline
\end{tabular}
\end{center}
\label{tab:param}
\end{table}

Since the pressure in the higher permeability fracture is established very fast, we will assume it to be constant along the fracture and equal to the confining stress $\sigma$ in the caprock. Therefore, when considering the pressure diffusion problem, the fracture will be represented as a boundary condition for the pressure 
\begin{equation}
\label{eq:bound-0}
\left. p(x,y,t) \right|_{|x| \le w/2, y = 0} = \sigma.
\end{equation}
Another boundary condition is
\begin{equation}
\label{eq:bound-inf}
\left. p(x,y,t) \right|_{x=\pm\infty} = p_0,
\end{equation}
where $p_0$ is the initial pressure in the aquifer, $p_0 > \sigma$. We note that $p_0$ is noticeably higher than the fluid pressure value before the CO$_2$ injection. Within these injection years high pressure propagates from the injection well in the aquifer, and we assume that far from the fracture the pressure remains constant. Then we assume no flux outside the aquifer, except for in the fracture
\begin{equation}
\label{eq:noflux-top}
\left. \frac{\partial p(x,y,t)}{\partial y} \right|_{|x| > w/2, y = 0} = 0
\end{equation}
and
\begin{equation}
\label{eq:noflux-bot}
\left. \frac{\partial p(x,y,t)}{\partial y} \right|_{y = h} = 0,
\end{equation}
where $h$ is the thickness of the aquifer (see Figure \ref{fig:scheme}).

For further consideration it is convenient to rewrite the problem in terms of dimensionless variables. We use the thickness of the aquifer $h$ as a unit of length and $h^2/c_f$ as a unit of time. Therefore, we introduce the following dimensionless variables:
\begin{equation}
\label{eq:dimensionless}
\chi \equiv x/h ~~~~~~ \xi \equiv y/h ~~~~~~ \tau \equiv t c_f/h^2.
\end{equation}
Then we consider the dimensionless pressure
\begin{equation}
\label{eq:f}
f(\chi,\xi,t) \equiv \frac{p(x,y,t) - \sigma}{p_0 - \sigma}.
\end{equation}
Therefore, the diffusion problem can be rewritten as:
\begin{subeqnarray}
\label{eq:diff2d-dim}
\frac{\partial f(\chi,\xi,\tau)}{\partial \tau} = \frac{\partial^2 f(\chi,\xi,\tau)}{\partial \chi^2} + \frac{\partial^2 f(\chi,\xi,\tau)}{\partial \xi^2} \\
\slabel{eq:bound-0-dim}
\left. f(\chi,\xi,\tau) \right|_{|\chi| \le w/2h, \xi = 0} = 0, \\
\slabel{eq:bound-inf-dim} 
\left. f(\chi,\xi,\tau) \right|_{\chi=\pm\infty} = 1, \\
\slabel{eq:noflux-top-dim}
\left. \frac{\partial f(\chi,\xi,\tau)}{\partial \xi} \right|_{|\chi| > w/2h, \xi = 0} = 0, \\
\slabel{eq:noflux-bot-dim}
\left. \frac{\partial f(\chi,\xi,\tau)}{\partial \xi} \right|_{\xi = 1} = 0.
\end{subeqnarray}

\subsection{Two-Dimensional Model}
\label{2D}

\begin{figure}[h]
  \centering
  \includegraphics[width=1.0\linewidth]{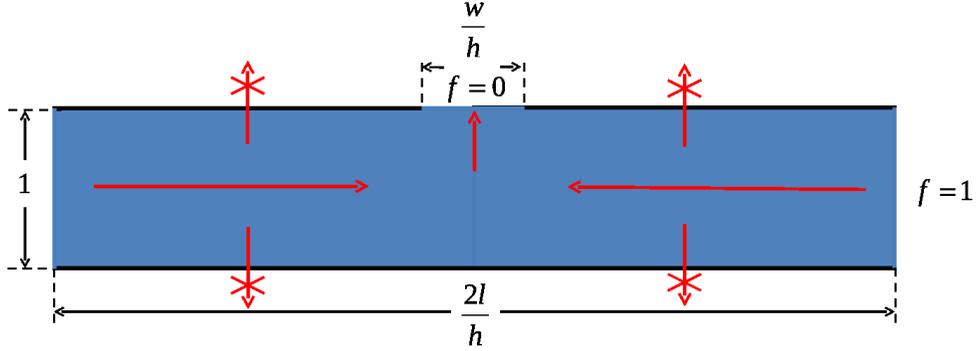}  
  \caption{Two-dimensional model of the aquifer used for numerical solution (not to scale). The dimensionless pressure $f$ at the boundaries $\chi = \pm \ell$ is $1$, the pressure at the fracture opening is $0$, and no-flux conditions are prescribed at the upper and lower boundaries.}
  \label{fig:2D-full}
\end{figure}

An analytical solution of the mixed boundary-value problem \ref{eq:diff2d-dim} formulated above is non-trivial and therefore we will solve it numerically. In the initial dimensional formulation we have two characteristic length scales: the fracture aperture $w \sim 10^{-6} - 10^{-2}$~m, and the thickness of the aquifer $h = 20$~m. Hence $h \gg w$. When solving the problem numerically one more length is introduced -- the length of the domain $\ell$ (for the half of the problem, i.e. $\chi > 0$). In order to model an infinite domain, we need $\ell \gg h$, which is typical of the actual physical problem. The schematic for this problem is shown in the Figure \ref{fig:2D-full}.

Therefore, in terms of dimensionless variables we have the following strong inequalities for the characteristic lengths:
\begin{equation}
\frac{w}{h} \ll 1 \ll \frac{\ell}{h}.
\end{equation}
The dimensionless times corresponding to these three lengths are 
\begin{equation}
\tau_w \equiv \frac{w^2}{h^2} ~~~~~ \tau_h \equiv 1 ~~~~~ \tau_{\ell} \equiv \frac{\ell^2}{h^2},
\end{equation}
so that 
\begin{equation}
\tau_w \ll \tau_h \ll \tau_{\ell}.
\end{equation}
Below we will present the results of a numerical solution of the problem and see how it differs on different time scales. 

For the sake of simplicity we take $\frac{w}{h} = \frac{1}{15}$ and $\frac{\ell}{h} = 30$, which will give us characteristic times $\tau_w = 1/225$ and $\tau_{\ell} = 900$. We solve the problem numerically on a rectangular mesh with 400 $\times$ 100 nodes; the mesh is refined near the fracture opening in both horizontal and vertical directions. The numerical solution is performed using Dynaflow -- a nonlinear transient finite element analysis program \cite{dynaflow}.

\subsection{One-Dimensional Model}
\label{1D}

Since the thickness of the aquifer $h$ is much smaller then its length, the flux in the aquifer is mostly in one horizontal direction (except in the vicinity of the fracture), see Figure \ref{fig:scheme}. Due to the symmetry, the flux $Q$ in this problem is twice the flux in the semi-infinite one-dimensional (1D) problem 
\begin{subeqnarray}
\label{eq:diff1d-dim}
\frac{\partial f(\chi,\tau)}{\partial \tau} = \frac{\partial^2 f(\chi,\tau)}{\partial \chi^2} \\
\slabel{eq:chi-0}
\left. f(\chi,\tau) \right|_{\chi = 0} = 0, \\
\slabel{eq:chi-inf} 
\left. f(\chi,\tau) \right|_{\chi=\infty} = 1.
\end{subeqnarray}
The analytical solution for this 1D diffusion problem is given by \cite{Carslaw-Jaeger}
\begin{equation}
\label{eq:f(y,t)}
f(\xi,t) = \mathrm{erf}\left[\frac{\chi}{2 \sqrt{\tau}} \right],
\end{equation}
where $\mathrm{erf}(x)$ is the error function. Calculating the partial derivative of the dimensionless pressure $f$ using the solution (\ref{eq:f(y,t)}) we obtain 
\begin{equation}
\label{eq:partial-f}
\frac{\partial f(\chi,\tau)}{\partial \chi} = \frac{1}{\sqrt{\pi \tau}} \exp\left[- \frac{\chi^2}{4 \tau} \right],
\end{equation}
and, therefore,
\begin{equation}
\label{eq:partial0-f}
\left. \frac{\partial f(\chi,\tau)}{\partial \chi} \right|_{\chi=0+}  = \frac{1}{\sqrt{\pi \tau}}.
\end{equation}
Darcy's law for the 1D case gives 
\begin{equation}
\label{eq:Darcy}
Q_{1D} = - \frac{h z k}{\mu} \left. \frac{\partial p(x,t)}{\partial x} \right|_{x=0+},
\end{equation}
where $z$ is the dimension of the aquifer perpendicular to the plane of Figure \ref{fig:scheme}, and $h z$ is the aquifer cross-sectional area. Therefore
\begin{equation}
\label{eq:Q1D}
Q_{1D} = \frac{z k}{\mu} \frac{(p_0 - \sigma)}{\sqrt{\pi \tau}} = \frac{h z k}{\mu} \frac{(p_0 - \sigma)}{\sqrt{\pi c_f t}}.
\end{equation}
Also, it has to be noted that $Q(t) \simeq 2 \times Q_{1D}(t)$. This coefficient 2 reflects that in the 2D problem  the fluid is coming from both $x = \pm \infty$ in Figure \ref{fig:scheme}. 

\section{Results}
\label{results}

\subsection{Calculation of the Flux through the Opening}

\begin{figure}[h]
  \centering
  \includegraphics[width=1.0\linewidth]{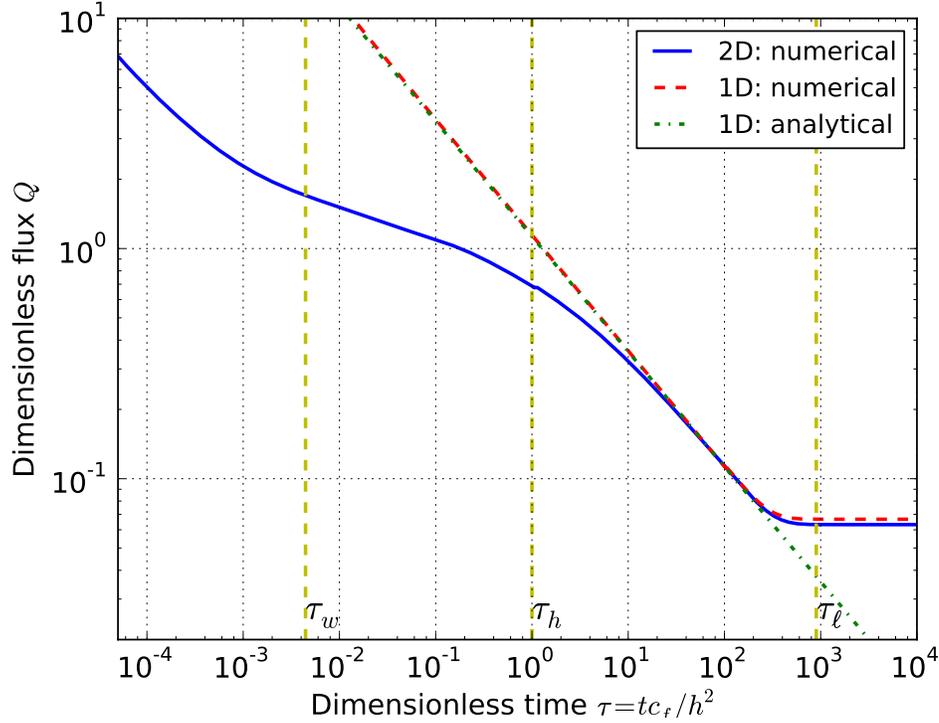}  
  \caption{The flux $Q$ normalized by $zk(p_0-\sigma)/\mu$ as function of dimensionless time $\tau = t c_f/h^2$ (log-log scale). The solid line (blue) represents the result of numerical solution for the 2D domain. The dashed line (red) represents the numerical solution ($\times 2$) of the problem for the 1D domain of the length $\ell$. The dash-dot line (green) gives the analytical solution of the non-steady 1D problem on the semi-infinite domain, see Eq. (\ref{eq:Q1D}). The vertical dashed lines represent the characteristic times $\tau_w$, $\tau_h$ and $\tau_{\ell}$.}
  \label{fig:stages}
\end{figure}

Two series of numerical simulations were carried out, with 1000 time steps each. The time steps were made a geometric sequence with the common ratio $1.1$. The first series started with the time step $10^{-5}$ and finished at the time $\sim 1$. The second series started with the time step $10^{-1}$ and finished at the time  $\sim 10^4$. Figure \ref{fig:stages} shows the flux $Q$ as a function of time. Figure \ref{fig:stages} clearly reveals four different stages of time evolution of the flux.

Stage 1: Fracture size effects for $\tau \ll \tau_w \simeq 4 \times 10^{-3}$ -- this period covers initial times when the diffusion perturbation spreads just in the vicinity of the fracture opening (length scale $w/h$).

Stage 2: 2D (aquifer thickness) effects for $\tau_w \ll \tau \ll \tau_h \simeq 1$ -- the vertical profile of the pressure is being established and the diffusion perturbation spreads towards the bottom of the aquifer (length scale $1$).

Stage 3: 2D effects vanish for $\tau_h \ll \tau \ll \tau_{\ell} \simeq 10^{3}$ -- the vertical profile of the pressure is established and the diffusion process is effectively 1D. We expect $Q(\tau) \propto \tau^{-1/2}$ [Eq. (\ref{eq:Q1D})]; this asymptotic behavior is clearly seen from the analytical solution of the 1D problem and agreement with the numerical simulations for the 1D problem.

Stage 4: Finite domain effects for $\tau_{\ell} \simeq 10^{3} \ll \tau$ -- this stage is a consequence of replacing a semi-infinite domain with the length $\ell$. At times $\gg \tau_{\ell}$ the steady state diffusion profile is established in the whole domain and, therefore, the flux $Q(\tau) = const \propto 1/\ell$ for numerical solutions of both 1D and 2D problems.

Thus, we conclude that for realistically long aquifers (kilometers), the solution of the 2D problem can be reasonably approximated by twice the solution of a 1D problem starting from Stage 3, i.e. at times $\tau \gg \tau_h$. For the typical situation (e.g. In Salah) $h = 20$~m, $c_f = 0.215$~m$^2$/s, therefore  the latter strong inequality is equivalent to $t \gg t_h = h^2/c_f \simeq 2000$~s, i.e. the 1D approximation works after $t \gg $ 35 minutes. 

\begin{figure}[h]
  \centering
  \includegraphics[width=1.0\linewidth]{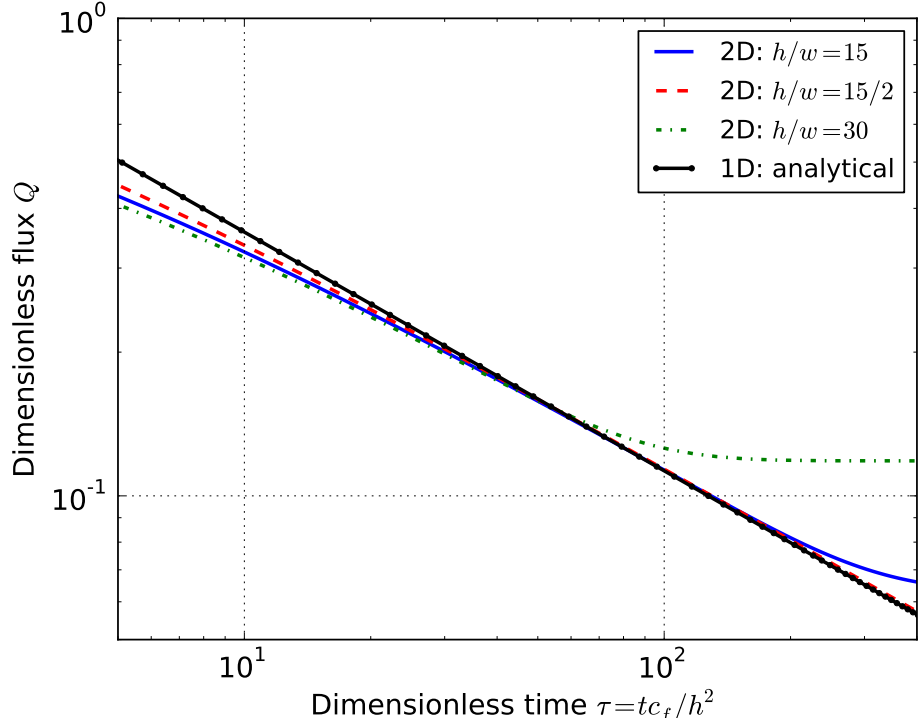}  
  \caption{The flux $Q$ normalized by $zk(p_0-\sigma)/\mu$ as a function of time calculated from the numerical solution of the 2D problem for three different thicknesses of the aquifer. Time scales for each curve corresponds to the thickness.}
  \label{fig:hs}
\end{figure}

The one-dimensional solution for the flux Eq. (\ref{eq:Q1D}) is proportional to the thickness of the aquifer. Let us see whether this is the case for the numerical solution of the 2D problem. We plot the results for 2D fluxes for three different thicknesses, plotting each curve with the corresponding time scale, i.e. for thickness $h$ with scale $\tau = tc_f/h^2$, for $h/2$ with scale $4tc_f/h^2$ and $2h$ with scale $tc_f/4h^2$. Figure \ref{fig:hs} shows that with such time scales the curves coincide. These results also show that for sufficiently long time the flux $Q$ for the 2D problem does not depend on the size of the opening (the fracture aperture). This result is in line with the analytical expression for the 1D case Eq. (\ref{eq:Q1D}).

\subsection{Solution for the Fracture Length}

The material balance Eq. (\ref{eq:balance}) provides the relation between the flux $Q$ (discharge from the aquifer) and the volume of the fracture. The volume of an elliptical fracture is given by
\begin{equation}
\label{eq:volume}
V(t) = \frac{\pi}{4} w(t) L(t) z.
\end{equation}
Using Darcy's law [Eq. (\ref{eq:Darcy})] and (\ref{eq:volume}) in Eq. (\ref{eq:balance}), we obtain:
\begin{equation}
\label{eq:balance2}
z \frac{\pi}{4} \frac{d}{dt} \left[ w(t) L(t) \right] = 2 h z \frac{k}{\mu} \left. \frac{\partial p(x,t)}{\partial x} \right|_{x=0+}.
\end{equation}
Note that $z$ cancels in this equation.

We assume that within all of the time from fracture initiation to time $t$, the pressure evolution is governed by a 1D diffusion problem. We also assume that the initial volume of the fracture at $t = 0$ is small; then we can rewrite Eq. (\ref{eq:balance2}) in the integral form 
\begin{equation}
\label{eq:balance3}
w(t) L(t) = \frac{8}{\pi} \frac{h k}{\mu} \int\limits_{0}^{t} \left. \frac{\partial p(x,t')}{\partial x} \right|_{x=0+} dt'.
\end{equation}
Substituting Eq. (\ref{eq:Q1D}) into Eq. (\ref{eq:balance3}) we get
\begin{equation}
\label{eq:balance4}
w(t) L(t) = \frac{8}{\pi^{3/2}} \frac{h k (p_0 - \sigma) }{\mu c_f^{3/2}} \int\limits_{0}^{t} {t'}^{-1/2} dt'.
\end{equation}
Integrating Eq. (\ref{eq:balance4}) we have
\begin{equation}
\label{eq:balance5}
w(t) L(t) = \frac{16}{\pi^{3/2}} \frac{h k (p_0 - \sigma)}{\mu c_f^{1/2}} t^{1/2}.
\end{equation}

In order to obtain an explicit expression for $L(t)$ we need to substitute the formula for $w(t)$ as a function of $L$ and $Q$, derived by Geertsma and de~Klerk \cite{Geertsma-deKlerk}
\begin{equation}
\label{eq:GdK}
w \simeq 2.1 \left[ \frac{\mu}{G} \frac{Q}{z} L^2 \right]^{1/4}, 
\end{equation}
where $G$ is the shear modulus of the rock, $G = \frac{E}{2(1+\nu)}$, and $E$ and $\nu$ are, respectively, Young's modulus and Poisson's ratio of the caprock. Eq. (\ref{eq:GdK}) was derived for hydraulic fracture, when the discharge $Q$ is controlled by the operator. In our case discharge $Q(t)$ is determined by the material balance equation and fracture parameters. Therefore, substituting Eq. (\ref{eq:Q1D}) into Eq. (\ref{eq:GdK}), taking into account Eq. (\ref{eq:dimensionless}), we find
\begin{equation}
\label{eq:GdK-time}
w(t) = 2.1 \left[ \frac{2}{\pi^{1/2}} \frac{h k (p_0 - \sigma)}{G c_f^{1/2} t^{1/2}} L(t)^2 \right]^{1/4}.
\end{equation}

\begin{figure}[h]
  \centering
  \includegraphics[width=1.0\linewidth]{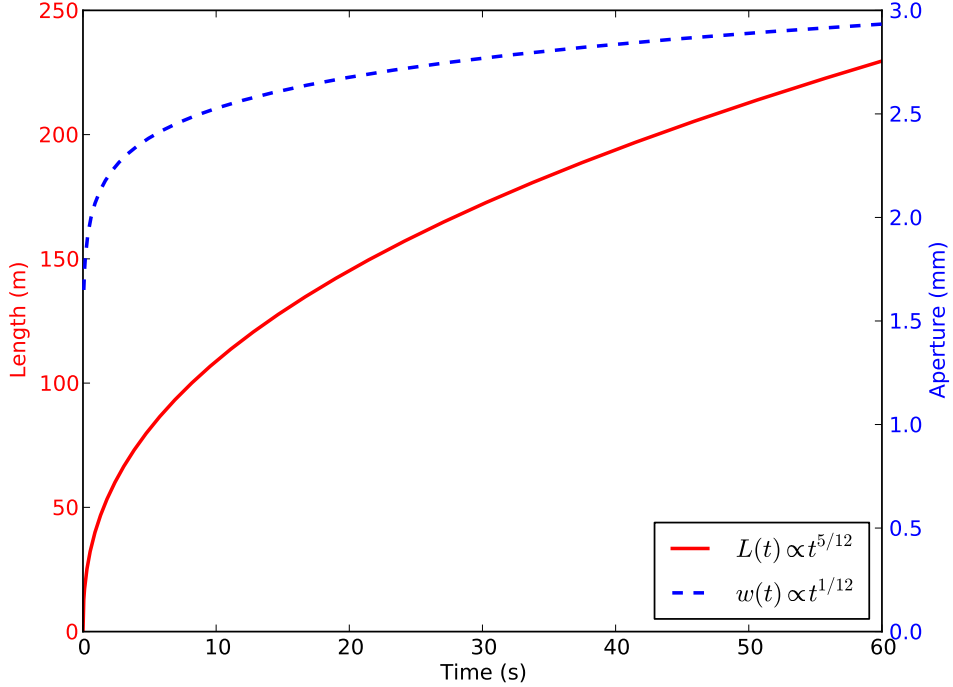}  
  \caption{Predictions for the fracture length and aperture evolution according to Eqs. Eq. (\ref{eq:GdK-time}), (\ref{eq:L-beta}) and (\ref{eq:beta}) for the physical parameters from Table \ref{tab:param}. The parameters of the mixture were calculated using the average weighted with the saturations.}
  \label{fig:evolution}
\end{figure}

Substituting Eq. (\ref{eq:GdK-time}) into Eq. (\ref{eq:balance5}) we arrive at the length of the fracture $L(t)$
\begin{equation}
\label{eq:L1}
L^{3/2}(t) = \frac{16}{2.1 \cdot 2^{1/4} \pi^{11/8}} \frac{k^{3/4} h^{3/4} G^{1/4} (p_0 - \sigma)^{3/4}}{\mu c_f^{3/8}} t^{5/8}
\end{equation}
or finally
\begin{equation}
\label{eq:L-beta}
L(t) = \beta t^{5/12}
\end{equation}
where
\begin{equation}
\label{eq:beta}
\beta \equiv 1.2 \frac{k^{1/2} h^{1/2} G^{1/6} (p_0 -\sigma)^{1/2}}{\mu^{2/3} c_f^{1/4}}.
\end{equation}
Characteristic values for our problem are: $k = 5 \times 10^{-14}$~m$^2$, $G = 8.7$~GPa, $\mu = 1.425 \times 10^{-4}$~Pa $\cdot$ s, $c_f = 0.215$~m$^2$/s, $p_0 - \sigma = 2$~MPa (see Table \ref{tab:param}). These estimates give us $\beta = 42$~m/s$^{-5/12}$. Substituting Eqs. (\ref{eq:L-beta}) and (\ref{eq:beta}) into Eq. (\ref{eq:GdK-time}), yields the time dependence of the fracture aperture. The evolution of the fracture length and aperture are shown in Figure \ref{fig:evolution}. We note that initially the fracture propagation is very fast: $100$ meters within less then a minute after initiation. However, such a rate is similar to the rate of propagation of hydraulic fractures \cite{Geertsma-deKlerk}.

\subsection{Correction for the Depth}
\label{sec:corrections}

Although in our schematic in Figure \ref{fig:scheme} the fracture is vertical, the solution we derive is applicable to a fracture propagating in any direction. In the current subsection we consider a vertical fracture only.

The driving force for fracture propagation is the difference between the fluid pressure in the fracture and the confining total horizontal stress in the caprock $(p_0 - \sigma)$. In the analytical solution for the fracture length Eq. (\ref{eq:L-beta}) $(p_0 - \sigma)$ is assumed constant, i.e. we assume that neither the confining stress nor hydrostatic pressure change with depth. Both effects can be important when the fracture propagates large enough distances toward the earth surface. Let us estimate how these values vary with the depth, i.e. consider $p = p(y)$ and $\sigma_H = \sigma_H(y)$. The values of $y$ are negative for the considerations below (Fig. \ref{fig:scheme}).

The total horizontal stress in the caprock $\sigma_H(y)$ is by definition the sum of the effective horizontal stress $\sigma_H'(y)$ and the water pressure $p_w(y)$ in the caprock
\begin{equation}
\label{eq:sigma_H}
\sigma_H(y) = \sigma_H'(y) + p_w(y).
\end{equation}
Here we use the sign convention for soil mechanics: compressive stresses have positive values. The caprock is saturated with brine (water) and CO$_2$ from the aquifer does not enter it; also the pressure in the caprock is not perturbed by the high pressure in the aquifer due to low permeability of shale. The effective horizontal stress is related to the effective vertical stress $\sigma_V'(y)$ through the lateral stress coefficient $K_{H}$
\begin{equation}
\label{eq:sigma_H'}
\sigma_H'(y) = K_{H} \sigma_V'(y).
\end{equation}
The vertical effective stress can then be calculated easily:
\begin{equation}
\label{eq:sigma_V'}
\sigma_V'(y) = \sigma_V'(0) + (1 - \phi_s) (\rho_s - \rho_w) g y,
\end{equation}
where $\sigma_V'(0)$ corresponds to the caprock-aquifer boundary, $\phi_s$ and $\rho_s$ are the porosity and density of the caprock respectively, $\rho_w$ is the density of the brine (water), and $g$ is the gravitational acceleration. Evidently, according to Eq. (\ref{eq:sigma_V'}) the effective vertical stress decreases with elevation, since the possible values of $y$ are negative.

The fluid pressure in the fracture at a certain height $y$ is given by
\begin{equation}
\label{eq:p_f}
p_f(y) = p_f(0) + \rho_f g y,
\end{equation}
where $\rho_f$ is the fluid density, calculated in accordance with the value of residual saturation (Table \ref{tab:param}). The water pressure in the caprock also changes with the depth
\begin{equation}
\label{eq:p_w}
p_w(y) = p_w(0) + \rho_w g y .
\end{equation}
Finally, collecting Eqs. (\ref{eq:sigma_H}) -- (\ref{eq:p_w}), we obtain the dependence of the driving force on the depth 
\begin{equation}
\label{eq:driv}
p_f(y) - \sigma_H(y) = p_f(0) - \sigma_H(0) - \left[ K_H (1 - \phi_s) (\rho_s - \rho_w) + (\rho_f -\rho_w)  \right] g y,
\end{equation}
where the last term is a positive value, increasing with elevation (due to $y < 0$). Substituting the values of the physical parameters for our system, and using $K_H = 0.46$ \cite{Gor-thermal}, calculated based on in situ stresses reported in \cite{Morris-2011}, we obtain the increase of the driving force per 1 meter of decrease of the depth, i.e.
\begin{equation}
\label{eq:elevat}
p_f(y) - \sigma_H(y) = p_0 - \sigma - \alpha y,
\end{equation}
where 
\begin{equation}
\label{eq:alpha}
\alpha \equiv \left[ K_H (1 - \phi_s) (\rho_s - \rho_w) + (\rho_f -\rho_w)  \right] g = 7.1 ~ \mathrm{kPa/m}
\end{equation}

The driving force for the fracture propagation at the fracture tip is determined by Eq. (\ref{eq:elevat}) with $y = -L(t)$. Therefore the analytical solution of the diffusion problem cannot be readily modified to take the hydrostatic and geostatic effects into account. Eq. (\ref{eq:elevat}) shows that the rate of the fracture propagation increases monotonically and our prediction for the rate of propagation is an estimate from below. 

\subsection{Pressure Diffusion in the Fracture}
\label{sec:flow}

We assumed that the pressure in the fracture is established instantaneously. The pressure evolution is determined by the ``diffusion coefficient'', Eq. (\ref{eq:c_f}), which is proportional to the permeability. Thus, in order to assume the diffusion in the fracture is fast compared to that in the aquifer, we need the permeability of the aquifer $k$ to be much lower than the permeability $\tilde{k}$ inside the fracture, i.e.
\begin{equation}
\label{eq:perm}
k \ll \tilde{k}.
\end{equation}
When the fracture propagates, its aperture $w(t)$ increases in time according to Eq. (\ref{eq:GdK-time}). The increase of the aperture causes the increase of permeability $\tilde{k} = \tilde{k}(t) \simeq w^2(t)/12$ \cite{Snow-1969}. However, this change makes the strong inequality (\ref{eq:perm}) even stronger. For $k = 50$~mD the strong inequality (\ref{eq:perm}) is valid when the fracture aperture $w \geq 2.5 \times 10^{-6}$~m. Using the parameters from Table \ref{tab:param}, our estimates for the initial fracture aperture based on the work of ref. \cite{Barenblatt-1962} give the initial fracture aperture $w(0) \geq 10^{-5}$~m, so the strong inequality (\ref{eq:perm}) is fulfilled starting already from the fracture initiation. 

\section{Conclusion}
\label{conclusion}

The safety of CO$_2$ storage in deep saline aquifers relies on the integrity of the caprock; fractures in the caprock may serve as pathways for CO$_2$ leakage if they propagate long enough. In this paper we present a theoretical model for propagation of a fracture driven by fluid outflow from a low-permeability aquifer. Since the pressure in the fracture is established very fast, the outflow is governed by the slower process -- pressure diffusion in the aquifer. By solving the 2D problem numerically, we show that after a relatively short time it can be approximated by the solution of 1D diffusion problem. The latter is solved numerically and analytically. 

Based on our solution of the diffusion problem, and the relation for the fracture geometry derived in the hydraulic fracture literature, we derive an analytical expression for the fracture propagation length as a function of time. Our simple model can be used together with the results of geomechanical simulations when the explicit consideration of fracture propagation is not included. This approach provides an estimate to the rate of fracture propagation based on the results of continuum mechanics simulations, without involving laborious simulations of fracture propagation.

Using the geomechanical and material parameters for the aquifer at In Salah, we predict the length of a hypothetical fracture propagation to be of the order of a hundred of meters within the first minute after initiation. This rate is extremely fat and is close to the typical rates of propagation of hydraulic fractures \cite{Geertsma-deKlerk}. 

We also estimate the depth correction to the driving force for the fracture propagation. We show that the changes of confining horizontal stress and hydrostatic pressure with elevation lead to an additional increase of the driving force for fracture propagation of the order of $7$ kPa per meter of elevation. Therefore our estimate for the rate of the fracture propagation is an estimate from below. Besides the In Salah site, the proposed model is also applicable to a number of aquifers currently used for CO$_2$ storage. As such, the following sites have aquifers with low permeability ($\sim 1-10$~mD): Nagaoka (Japan), Alberta Basin (Canada), MRCSP Michigan Basin (USA), Gorgon (Australia) \cite{Michael-2009}.

Fracturing of the caprock can still be a serious safety concern. In order to arrest the fracture, the fluid pressure must be decreased. Shutting down the injection will not have an immediate effect. After several years of the continuous injection the pressure in the reservoir is spread over several kilometers. Therefore, even if injection is stopped, the aquifer will remain over-pressured for a long time.

\section*{Acknowledgements}
Funding for this research has been provided by the Carbon Mitigation Initiative (http://cmi.princeton.edu) sponsored by BP. We thank George Scherer for fruitful discussions and Allyson Sgro for useful comments on the manuscript. We also thank the anonymous reviewers for the constructive comments that led to significant improvements in the manuscript.


\begin{thebibliography}{10}
\expandafter\ifx\csname url\endcsname\relax
  \def\url#1{\texttt{#1}}\fi
\expandafter\ifx\csname urlprefix\endcsname\relax\def\urlprefix{URL }\fi

\bibitem{Cappa-Rutqvist}
F.~Cappa, J.~Rutqvist, Modeling of coupled deformation and permeability
  evolution during fault reactivation induced by deep underground injection of
  {CO}$_2$, International Journal of Greenhouse Gas Control 5~(2) (2011) 336 --
  346.

\bibitem{Iding-2010}
M.~Iding, P.~Ringrose, Evaluating the impact of fractures on the performance of
  the {In Salah} {CO}$_2$ storage site, International Journal of Greenhouse Gas
  Control 4~(2) (2010) 242 -- 248.

\bibitem{Luo-2010}
Z.~Luo, S.~L. Bryant, Influence of thermo-elastic stress on {CO}$_2$ injection
  induced fractures during storage, in: SPE International Conference on
  {CO}$_2$ Capture, Storage, and Utilization, 10-12 November 2010, New Orleans,
  Louisiana, USA, 2010.

\bibitem{Luo-2011}
Z.~Luo, S.~L. Bryant, Influence of thermoelastic stress on fracturing a
  horizontal injector during geological {CO}$_2$ storage, in: Canadian
  Unconventional Resources Conference, 15-17 November 2011, Alberta, Canada,
  2011.

\bibitem{Preisig-Prevost}
M.~Preisig, J.~H. Pr\'evost, Coupled multi-phase thermo-poromechanical effects.
  case study: {CO}$_2$ injection at {In Salah}, {Algeria}, International
  Journal of Greenhouse Gas Control 5~(4) (2011) 1055 -- 1064.

\bibitem{Goodarzi-2012}
S.~Goodarzi, A.~Settari, D.~Keith, Geomechanical modeling for {CO}$_2$ storage
  in nisku aquifer in wabamun lake area in canada, International Journal of
  Greenhouse Gas Control 10 (2012) 113 -- 122.

\bibitem{Gor-thermal}
G.~Y. Gor, T.~R. Elliot, J.~H. Pr\'evost, Effects of thermal stresses on
  caprock integrity during {CO}$_2$ storage, International Journal of
  Greenhouse Gas Control 12 (2013) 300--309.

\bibitem{Humez-2011}
P.~Humez, P.~Audigane, J.~Lions, C.~Chiaberge, G.~Bellenfant, Modeling of
  {CO}$_2$ leakage up through an abandoned well from deep saline aquifer to
  shallow fresh groundwaters, Transport in Porous Media 90~(1) (2011) 153--181.

\bibitem{Smith-2011}
J.~Smith, S.~Durucan, A.~Korre, J.-Q. Shi, Carbon dioxide storage risk
  assessment: Analysis of caprock fracture network connectivity, International
  Journal of Greenhouse Gas Control 5~(2) (2011) 226--240.

\bibitem{Geertsma-deKlerk}
J.~Geertsma, F.~de~Klerk, A rapid method of predicting width and extent of
  hydraulically induced fractures, Journal of Petroleum Technology 21~(12)
  (1969) 1571--1581.

\bibitem{Khr-Zheltov}
S.~A. Khristianovich, Y.~P. Zheltov, Formation of vertical fractures by means
  of highly viscous liquid, Proc., Fourth World Pet. Cong., Rome Sec. II (1955)
  579--86.

\bibitem{Rutqvist-2010}
J.~Rutqvist, D.~W. Vasco, L.~Myer, Coupled reservoir-geomechanical analysis of
  {CO}$_2$ injection and ground deformations at {In Salah}, {Algeria},
  International Journal of Greenhouse Gas Control 4~(2) (2010) 225 -- 230.

\bibitem{Barenblatt-1962}
G.~I. Barenblatt, The mathematical theory of equilibrium cracks in brittle
  fracture, Vol.~7 of Advances in Applied Mechanics, Elsevier, 1962, pp. 55 --
  129.

\bibitem{Detournay-2004}
E.~Detournay, Propagation regimes of fluid-driven fractures in impermeable
  rocks, International Journal of Geomechanics 4~(1) (2004) 35--45.

\bibitem{Nilson-1988}
R.~H. Nilson, Similarity solutions for wedge-shaped hydraulic fractures driven
  into a permeable medium by a constant inlet pressure, International Journal
  for Numerical and Analytical Methods in Geomechanics 12~(5) (1988) 477--495.

\bibitem{Zhang-Scherer}
J.~Zhang, G.~W. Scherer, A novel method for measuring permeability of shale,
  International Journal of Rock Mechanics and Mining Sciences 53 (2012)
  179--191.

\bibitem{Coussy}
O.~Coussy, Poromechanics, 2nd Edition, Wiley, 2004.

\bibitem{dynaflow}
J.~H. Pr\'evost, {DYNAFLOW: A Nonlinear Transient Finite Element Analysis
  Program. Department of Civil and Environmental Engineering, Princeton
  University, Princeton, NJ (1981)}. http://blogs.princeton.edu/prevost/dynaflow/ (last update
  2013).

\bibitem{Carslaw-Jaeger}
H.~S. Carslaw, J.~C. Jaeger, Conduction of Heat in Solids (Second Edition),
  Oxford Science Publications, 1986.

\bibitem{Morris-2011}
J.~P. Morris, Y.~Hao, W.~Foxall, W.~McNab, A study of injection-induced
  mechanical deformation at the {In Salah} {CO}$_2$ storage project,
  International Journal of Greenhouse Gas Control 5~(2) (2011) 270 -- 280.

\bibitem{Snow-1969}
D.~T. Snow, Anisotropic permeability of fractured media, Water Resources
  Research 5~(6) (1969) 1273--1289.

\bibitem{Michael-2009}
K.~Michael, G.~Allinson, A.~Golab, S.~Sharma, V.~Shulakova, {CO}$_2$ storage in
  saline aquifers {II} - experience from existing storage operations, Energy
  Procedia 1~(1) (2009) 1973 -- 1980.

\end{thebibliography}
\end{document}